\documentclass[epj, final]{svjour}
\usepackage{latexsym}
\usepackage{graphicx}
\usepackage{hyperref}
\usepackage{cite}

\def\v#1{{\bf#1}}
\def\be{\begin{equation}}
\def\ee{\end{equation}}
\def\bea{\begin{eqnarray}}
\def\eea{\end{eqnarray}}

\def\fcal{\mbox{$\cal F\,$}}
\def\ccal{\mbox{$\cal C\,$}}
\def\dcal{\mbox{$\cal D\,$}}
\def\<{\langle}
\def\>{\rangle}

\begin{document}

\title{Fundamental constraints on two-time physics}
\author{E. Piceno\inst{1} \and A. Rosado\inst{2} \and E. Sadurn\'i\inst{2,3}}
\institute{Facultad de Ciencias F\'isico Matem\'aticas, Benem\'erita Universidad Aut\'onoma de Puebla, Apartado Postal 1152, Puebla, Puebla, M\'exico \and Instituto de F\'isica, Benem\'erita Universidad Aut\'onoma de Puebla,
Apartado Postal J-48, 72570 Puebla, M\'exico \and \emph{Electronic address:} sadurni@ifuap.buap.mx}
\date{\today}

\abstract{We show that generalizations of classical and quantum dynamics with two times lead to fundamentally constrained evolution. At the level of classical physics, Newton's second law is extended and exactly integrated in $1+2$ dimensional space, leading to effective single-time evolution for any initial condition. The cases $2+2$ and $3+2$ are also analyzed. In the domain of quantum mechanics, we follow strictly the hypothesis of probability conservation by extending the Heisenberg picture to unitary evolution with two times. As a result, the observability of two temporal axes is constrained by a generalized uncertainty relation involving level spacings, total duration of the effect and Planck's constant.}
\maketitle

\section{Introduction}

Two time physics has been a tantalizing possibility of dimensionally-extended descriptions of our world \cite{Scherk:1979zr, Rubakov:1983bz, Weinberg:1983xy, Alvarez:1983kt, Barrow:1987sr}, in particular for those extensions related to extra-dimensional high-energy physics and their inherent compactification mechanisms \cite{Hosotani:1983xw, Cremmer:1976zc, Flacke:2005hb, Kong:2005hn}. The focus of attention in such theoretical endeavours is the existence of phenomenological as well as consistency constraints that allow to answer the simplest, yet most profound question on the nature of space-time. Undoubtedly, all areas of physics are affected by its outcome. Noteworthy attempts \cite{Petriello:2002uu, Appelquist:2000nn, Haisch:2007vb, Goertz:2011hj, ADDM2002, DDG1999} have directed their efforts to distill numerical bounds on physical quantities -- i.e. energies or masses -- using a theoretical apparatus related to extended-metric versions of classical and quantum field theories. Posed in this language --i.e. a problem of group representations on dimensionally extended manifolds-- the challenge seems fit for particle physicists, who are well versed in mechanisms \cite{Giardino:2011zz, Allahverdi:2003aq} that either hide or expose measurable quantities in the domain of current experimental capabilities. We believe, however, that even more fundamental restrictions may arise from the point of view of dynamical systems when their evolution is extended to two parameters. At such an elementary level of formality, we encounter already a number of compromising questions.

Indeed, we may formulate initially symmetric two-time dynamics in compliance with Newton's second law and discover, with little effort, that any $1+2$ dimensional extension can be integrated under reasonable smoothness conditions on the functions involved. The main result is that the dynamics occur on a restricted surface and therefore only one time. This would be a surprising result in principle, since typical integrability conditions rest on the existence of integrals of the motion that match the system's dimensionality. Our results show that in the presence of two times, not only the integrability conditions must be rethought, but also that $1+2$ extensions render simple solutions with a preferred time axis at the level of classical mechanics. It is the resulting motion and not the shape of the equations what yields an effective one-time description.
We can formulate the $2+2$ and $3+2$ dimensional extensions in the same manner, therefore these extensions yield similar results. In these cases, two-time evolution takes place only in a dimensionally reduced surface of the force space. Thus we still have one preferred time direction for the case of a general force, and two times for very special cases.

Less restrictive conditions can be derived in quantum mechanical settings, either following the approach of dimensionally extended wave equations, or by extending the number of evolution generators that ensure unitary dynamics. In regard with the former, the conservation of probability is at stake: dimensionally extended continuity equations yield, upon integration, a conserved probability along a single time direction. In the present work we show that circumventing such a drawback via unitary evolution leads, after dynamical analysis, to classical limits constrained again by Newton's second law for two times --now at the level of quantum averages. However, our probability-conserving approach also shows that a full quantum-mechanical regime allows a small window of opportunity determined by a generalized uncertainty relation in energies and times. Furthermore, such a relation provides bounds on the observability of two-time physics in terms of fluctuations in level spacings, Planck's constant and the norm of elapsed times.

We present our discussion starting with the simplest classical example in section \ref{s1}, and we gradually move to an extended Heisenberg picture in section \ref{s2}. We conclude in section \ref{s3}.

\section{Fundamental aspects and motivation}

String theory and Kaluza-Klein compactification \cite{bailin1987} rely on dimensional extensions that are normally proposed in the space-like part of the Minkowski metric. Why not time? The old idea of introducing stationary modes for extra dimensions in mass operators (e.g. in dimensionally extended Klein-Gordon equations) does produce effective masses related to compactification scales. In the time-like part, these extensions would obviously enter with the opposite sign: if $\Box_{d+2}$ is the $d+2$-dimensional D'Alembert operator, one has that the wavefunction $\phi$ for a scalar particle of mass $m$ satisfies

\bea
\left[ \Box_{d+2} + \frac{m^2 c^2}{\hbar^2} \right] \phi = 0,
\label{add1}
\eea
where 
\bea
\Box_{d+2}= \frac{1}{c^2}\frac{\partial^2 }{\partial t_1^2} + \frac{1}{c^2}\frac{\partial^2 }{ \partial t_2^2}- \frac{\partial^2 }{ \partial x_1^2} - \cdots \frac{\partial^2 }{ \partial x_d^2}.
\label{add1.1}
\eea
The introduction of additional times would correct such a mass; if we set $\phi(\v x, t_1,t_2) = e^{-i\omega t_2}\tilde \phi(\v x, t_1)$, leads to

\bea
\left[ \Box_{d+1} + \frac{m^2 c^2}{\hbar^2} - \frac{\omega^2}{c^2} \right] \tilde \phi = 0.
\label{add2}
\eea
The new mass for this mode is then $\tilde m^2 c^4 = m^2 c^4 - \hbar^2 \omega^2$. Strong implications on the values of these masses would arise due to extra time axes. Usually, it is said \cite{bailin1987} that such extensions should not be considered due to an imaginary effective mass, but here we note that previously existing masses with the value $m$ (coming either from compactified space-like coordinates or other fundamental mechanisms) still allow a real positive $\tilde m$, up to a certain frequency cut off $\omega_{max}$ determined by $\tilde m^2 > 0$. This entails the existence of a fundamental period $\tau = 2\pi / \omega$ that should be compared with a fundamental length $R = \hbar/mc$. The condition that avoids tachyonic --imaginary mass-- representations of the Poincar\'e group acting on $d+1$-dimensional Minkowski space is the inequality $c \tau > R$, while the equality does not produce mass.

Other important observations on the existence of more times are related to the proper formulation of dynamical systems for two parameters, which is the main purpose of this paper. We admit that nature is described by integro-differential equations that govern physical quantities, e.g. the position of a particle in space, the value of a classical field, or the space dependence of probability densities in the quantum world. Regardless of the specific form of such equations, the evolution of a system is produced by nature once a set of initial data is provided. This should also apply for more than one time axis; therefore, a consistent description of this type of systems should be given. For this reason, and in the case of classical mechanics, it seems mandatory to address the problem from the point of view of Newton's second law, with a properly specified force and some initial conditions --in general, Hamiltonian systems of two times.   

In connection with experimental aspects, the situation must be described as purely observational. This is a common denominator in every extra-dimensional theory, both for space and time. One looks for quantites that can be produced or influenced by extra-dimensional physics --the typical example being the aforementioned masses-- and their values are detected in our $3+1$-dimensional world. Indeed, we shall see that this is not limited to the alluded masses, but also includes corrections to the time-energy uncertainty principle arising in quantum-mechanical transition probabilities and quantum mechanical fluctuations; specifically, in section \ref{S2.3} we shall analyze the standard deviation of a particle's position, leading to the conclusion that its value depends strongly on time-energy fluctuations satisfying a new inequality (\ref{27}). If the new inequality is confirmed, it would give a hint for new physics.

It should be clear though, that many effects could be falsified by other (new) physical considerations; the confirmation of a corrected mass or a new uncertainty principle is not an undeniable proof of extra dimensions, neither spatial nor temporal. Any experimental apparatus made in our world measures the effects produced in it, but the apparatus could hardly venture into other dimensions merely by its motion. In the same manner, we may touch our shadow projected on a wall, but our shadow cannot stick out its arm and shake our hand.

\section{Classical dynamics of two times \label{s1}}We start by allowing Newton's second law to accomodate two times via a function $x^î(t_1,t_2)$ representing position coordinates. Then, adopting units where mass can be ignored, there must be two differential and non-preferential equations of the form 

\bea
p^{i}_j = \frac{\partial x^{i}}{\partial t_j} + A^{i}_{j}, \quad i=1,2,\cdots,d \quad j=1,2
\label{1}
\eea
\bea
\frac{\partial p^{i}_{j}}{\partial t_k} = F^{i}_{kj}
\label{2}
\eea
where, for convenience, space indices are placed above quantities, and time indices below.
Eq. (\ref{1}) shows the necessity of considering two velocities and therefore two momenta, perhaps with a connection $A^{i}_j$ coming from gauge interactions. The partial derivatives with respect to times 1 and 2 are now required. Eq. (\ref{2}) expresses an extension of Newton's second law whenever $F^i_{jk}$ is a specified force -- now a tensor in time indices. As in any specified force governing the dynamics of a system, we have a generic dependence $F^i_{jk}(t_1,t_2; \v x(t_1,t_2))$. In addition, if $F$ depends explictly on the two times, there is little we could do to rule out the relevance of a second time, which compels to consider $F^i_{jk}(\v x(t_1,t_2))$ as a specific force law (autonomous differential equations). Now we may proceed to analyze the dynamical consequences of such a force field: combining (\ref{1}) and (\ref{2}) and abbreviating with $\partial_j$ the time derivatives, we obtain the following antisymmetrization in time indices

\bea
\partial_{\left[ k \right.} p^{i}_{\left. j\right]} = \partial_{\left[ k \right.} \partial_{\left. j\right]} x^{i} + \partial_{\left[ k \right.} A^{i}_{\left. j\right]}.
\label{3}
\eea
If $x^{i}$ is a sufficiently differentiable function (a reasonable requirement at least in some domain of the variables $t_1, t_2$) we may eliminate the first term and retain the antisymmetrization due to all other contributions to the momenta:

\bea
F^{i}_{\left[ jk\right]} = \partial_{\left[ j \right.} A^î_{\left. k \right]},
\label{4}
\eea
as expected. It is striking to see that any attempt at generalizing physics in this way is compromised by space dimensionality, as we now discuss.

\subsection{One spatial dimension}

 If $d=1$, the equations of motion and the chain rule can be used to prove the following relations

\bea
\partial_1 \partial_2 p_k = (p_1 - A_1) \frac{\partial \left( F_{k2} + \partial_{\left[ 2 \right.} A_{\left. k \right]} \right)}{\partial x},
\label{5}
\eea
\bea
\partial_2 \partial_1 p_k = (p_2 - A_2) \frac{\partial \left( F_{k1} + \partial_{\left[ 1 \right.} A_{\left. k \right]} \right)}{\partial x}.
\label{6}
\eea
As before, we admit that each function $p_k(t_1,t_2)$ is continuously differentiable and compare (\ref{5}) and (\ref{6}) for $k=1,2$

\bea
\frac{\partial F_{22}}{\partial x} (p_1-A_1) = \frac{\partial F_{12}}{\partial x} (p_2-A_2),
\label{7}
\eea
\bea
\frac{\partial F_{21}}{\partial x} (p_1-A_1) = \frac{\partial F_{11}}{\partial x} (p_2-A_2).
\label{8}
\eea
Surprisingly, we can establish now an equation for 'orbits' relating $p_1, p_2$ and the elements of the force tensor as functions of $x$. This is done by solving the previous system of equations to find

\bea
\left(\frac{p_1-A_1}{p_2-A_2}\right)^2 = \frac{F'_{11}F'_{12}}{F'_{21}F'_{22}} \equiv \Phi(x),
\label{9}
\eea 
\bea
F'_{11} F'_{22} = F'_{12} F'_{21},
\label{10}
\eea
\begin{figure*}
\begin{center}\includegraphics[width=8.6cm]{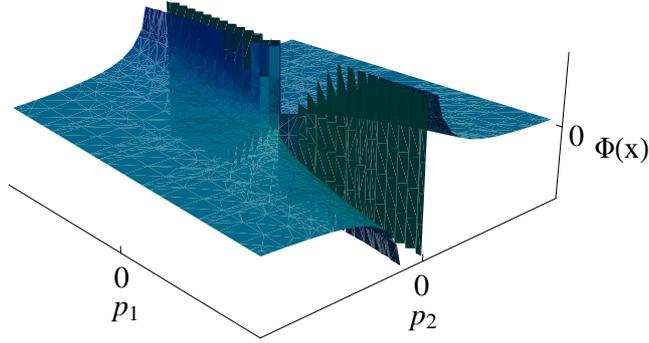}\end{center}
\caption{\label{fig:1} Dynamical surface of arbitrary two-time systems. $\Phi(x)$ is used as a coordinate instead of $x$ for better visibility.}
\end{figure*}
where we denote the derivative with respect to $x$ by a prime. These orbits are indeed surfaces that we describe in figure \ref{fig:1}. The surface is valid for any initial condition, excluding the possibility of a full foliation of the extended phase space $x,p_1,p_2$. This means that the initial value problem is well posed if the second velocity is constrained by the dynamics in the $(x,p_1)$ plane. Another way to look at this comes from a convenient definition of a field $\v \fcal =(\sqrt{F'_{22}F'_{21}},-\sqrt{F'_{11}F'_{12}})$, showing thus that all the evolution happens under the restriction

\bea
\v \fcal \cdot \nabla_{12}\, x(t_1,t_2) = 0.
\label{11}
\eea
As a consequence, the function $x(t_1,t_2)$ is constant in the direction parallel to $\fcal$. Regardless of the explicit form of $\fcal$, there is a family of non-intersecting curves in $(t_1,t_2)$ defining the evolution, which picks only one direction --but not the arrow of time-- as a result of consistent dynamics. It is also possible to find a restriction on the field $\fcal$ with the aim of supressing solutions with closed curves in the plane $t_1,t_2$ -- a most desirable condition, in view of causality. This type of analysis shall become important in the higher dimensional cases discussed below. 

\subsection{Two and three spatial coordinates}

The number of constraints emerging in the higher dimensional cases can be determined in a straightforward way. It is advantegeous to reproduce the one-dimensional calculation in (\ref{7}) and (\ref{8}) in order to find the corresponding dimensionally-extended expressions. The smoothnes condition on the functions $p^i_{j}$ (i.e. commutability of time derivatives) is expressed as

\bea
\partial_l F_{jk}^i = \partial_j F_{lk}^i.
\label{1.1}
\eea
Using the chain rule again and ignoring any explicit time dependence of $F$ (we are analyzing autonomous systems!), this equality acquires the form

\bea
\sum_{m=1,...,d}  \frac{\partial F_{jk}^i}{\partial x^m}  \partial_l x^m = \sum_{m=1,...,d} \frac{\partial F_{lk}^i}{\partial x^m}  \partial_j x^m
\label{1.2}
\eea
where $d$ is the spatial dimension. For simplicity, we consider now $A=0$ (symmetric $F$) and write (\ref{1.2}) as

\bea
\sum_{m=1,...,d} \left[ \frac{\partial F_{jk}^i}{\partial x^m}  p^m_l - \frac{\partial F_{lk}^i}{\partial x^m}  p^m_j  \right] = 0.
\label{1.3}
\eea
For $d=2$ this condition comprises in fact four equations that can be cast as a homogenous linear system with four velocities as variables:

\begin{equation}\label{1.4}
\left(\begin{array}{cccc}F^1_{21,x}&-F^1_{11,x}&F^1_{21,y}&-F^1_{11,y}\\
F^1_{22,x}&-F^1_{12,x}&F^1_{22,y}&-F^1_{12,y}\\
F^2_{21,x}&-F^2_{11,x}&F^2_{21,y}&-F^2_{11,y}\\
F^2_{22,x}&-F^2_{12,x}&F^2_{22,y}&-F^2_{12,y}\end{array}\right) \left( \begin{array}{c} p_1^1 \\ p_2^1 \\ p_1^2 \\ p_2^2 \end{array} \right)  =0.,
\end{equation}
where
\begin{equation}\label{defpar}
F_{jk,x^m}^i\equiv\frac{\partial F_{jk}^i}{\partial x^m} ,
\end{equation}
and $x\equiv x^1$, $y\equiv x^2$.
The analogue of (\ref{10}) now shows in the form of a determinant, i.e. any non-trivial force field allowing non-zero velocities must satisfy

\begin{equation}\label{1.5}
\det\left(\begin{array}{cccc}F^1_{21,x}&-F^1_{11,x}&F^1_{21,y}&-F^1_{11,y}\\
F^1_{22,x}&-F^1_{12,x}&F^1_{22,y}&-F^1_{12,y}\\
F^2_{21,x}&-F^2_{11,x}&F^2_{21,y}&-F^2_{11,y}\\
F^2_{22,x}&-F^2_{12,x}&F^2_{22,y}&-F^2_{12,y}\end{array}\right)=0. 
\end{equation}
With this condition and the general solution of (\ref{1.4}) we find once more a parallel restriction on the evolution of each coordinate:

\bea
\ccal \cdot \nabla_{12} x(t_1,t_2) = 0, \qquad
\dcal \cdot \nabla_{12} y(t_1,t_2) = 0
\label{1.6}
\eea
where the fields $\ccal$ and $\dcal$ are defined in terms of derivatives of the force, given in \ref{app1}. As noted in the previous section, more constraints can be imposed by means of physical considerations. For example, ruling out closed curves in $t_1, t_2$ for each function $x$ and $y$ entails the elimination of the curl of each parallel field

\bea
\nabla_{12} \times \ccal = 0, \qquad \nabla_{12} \times \dcal = 0.
\label{1.7}
\eea 
since each curl contains only one component, (\ref{1.7}) provides two additional restrictions. The number of conditions in our treatment beg for a careful count of restrictions and free variables. We may have effective one-time dynamics if the two vector fields $\ccal$, $\dcal$ turn out to be parallel; fortunately this constitutes a single restriction

\bea
\det \left( \begin{array}{cc} \ccal_1 & \ccal_2  \\  \dcal_1 & \dcal_2 \end{array}\right) = 0
\label{1.8}
\eea
which does not change the dimensionality of the original space. We have in principle 8 independent functions for the entries of the force tensor, but they reduce to 6 when symmetry is invoked. Eqns. (\ref{1.7}) reduce the dimensionality by two, but the most important condition for a non-trivial solution is (\ref{1.5}), which subtracts yet another dimension. In total, we are left with 3 free functions for a space of 8 components, and (\ref{1.8}) represents only a two-dimensional exclusion in such a space. In conclusion, non-trivial dynamics (i.e. with two times) is attainable for a reduced space of possible forces: in this sense, two-time physics must be driven by a special kind of force.

The three dimensional case leads to more intricate expressions, which however follow the same logic as in the previous discussion. A similar procedure using (\ref{1.3}) with $d=3$ leads to the restriction

\begin{equation} \det \left( \begin{array}{cccccc}
	F_{21,x}^1&-F_{11,x}^1&F_{21,y}^1&-F_{11,y}^1&F_{21,z}^1&-F_{11,z}^1\\
	F_{21,x}^2&-F_{11,x}^2&F_{21,y}^2&-F_{11,y}^2&F_{21,z}^2&-F_{11,z}^2\\
	F_{21,x}^3&-F_{11,x}^3&F_{21,y}^3&-F_{11,y}^3&F_{21,z}^3&-F_{11,z}^3\\
	F_{22,x}^1&-F_{12,x}^1&F_{22,y}^1&-F_{12,y}^1&F_{22,z}^1&-F_{12,z}^1\\
	F_{22,x}^2&-F_{12,x}^2&F_{22,y}^2&-F_{12,y}^2&F_{22,z}^2&-F_{12,z}^2\\
	F_{22,x}^3&-F_{12,x}^3&F_{22,y}^3&-F_{12,y}^3&F_{22,z}^3&-F_{12,z}^3\end{array} \right) =0.
\label{1.9}
\end{equation}
As before, the linear system can be expressed as a number of orthogonality conditions for some vector fields $\ccal_1$, $\ccal_2$ and $\ccal_3$ (the procedure to obtain their expressions is indicated in \ref{app2}):

\bea
\ccal_i \cdot \nabla_{12} x^{i} = 0, \quad \nabla_{12} \times \ccal_i = 0, \quad i=1,2,3.
\label{1.10}
\eea
Finally, the $3+2$ dimensional evolution is governed by tensors which are dimensionally restricted. We have $12$ original components, $3$ symmetry conditions, $3$ irrotational conditions,  and a vanishing determinant, i.e. $12-3-3-1=5$ dimensions. An accidental case inside such a space is still possible, since $\ccal_1 \, || \, \ccal_2 \, || \, \ccal_3$ yields effective one-time evolution, but this does not reduce the dimensionality any further. We reach thus the same conclusion as in the $d=2$ case.


\subsection{Remarks on classical results}

So far, we have found the necessary constraints {\it on the force fields and space dimensions\ }to produce variations of coordinates in two times. These constraints are conditions that should not be confused with the well-known theory of constraints proposed by Dirac \cite{Dirac1964}. However, there seems to be an important analogy with this interesting topic when dealing with special relativity of many bodies and it deserves a few comments. In this setting, multiple times emerge from various four-vectors $x_{\mu}^{(i)}$ for each particle, where $i$ is the particle index, and attached to each vector there is a seemingly independent time variable $x_{0}^{(i)}$. The constraints that emerge from this theory correspond to the invariance of each rest mass $m_i$, 
and they have the form $p^{(i)\mu}_ip^{(i)}_{\mu}=m_i^2c^2$, i.e. on-shell. 
Since each $p^{(i)0}_i$ is determined by momenta and masses through an equality, we are in the presence of a surface constraint. It must be stressed, however, that the apparent dependence on multiple times has been reduced in the past to single-time dynamics by Barut and coworkers \cite{barut}, as well as Moshinsky and Nikitin \cite{moshinsky1} and Moshinsky and Sadurni \cite{moshsadurni}, where it was possible to describe the dynamics of a composite system with a single Hamiltonian corresponding to the time-like component of the total four-momentum. Moreover, interactions between particles could be introduced using perpendicular (frame-dependent) four-vectors together with relative Jacobi coordinates generalized to Minkowski space -- here one should note the contrast with common criticisms related to the inconsistency of certain types of relativistic interactions \cite{lienert1}. Some nice results regarding this topic were summarized in \cite{moshinsky2}(in particular, Chapter XII, 315-316), especially those connected with the many-body Dirac oscillator. For more recent applications to spectroscopy of relativistic composites, see \cite{sadurni1, sadurni2}, where similar ideas came into play for strongly interacting relativistic particles. The treatment in \cite{sadurni1}, eqns. (6)-(11) is of special interest for this topic. In all, two-time physics of a single particle is not a particular case of typical relativistic multi-body physics, but it constitutes a parallel line of study for such problems. Furthermore, one should not deny the possibility that our result of a single surface in 3D phase space (\ref{10}), (\ref{11}) could be better formulated using Hamiltonian constraints. 

Also, a cautionary remark is in order. Although the bridge connecting the classical and the quantum world could be built based on classical constraints \cite{komar} through weak equalities and the Hilbert space obeying them, it is important to recognize that the quantization of classical constraints is not equivalent to the constraint of quantum systems. This fact was first noted by Jensen \cite{jensen}, further elaborated by daCosta \cite{dacosta} in connection with curvature, and it is of utmost importance to our endeavours. Very specific effects emerging from this difference can be found in \cite{sadurni3, sadurni4}, e.g. bound states with no classical counterpart.
For this reason and in the following sections, the quantum side of the two-time problem will be tackled from scratch; this will ensure generality regarding constraints.

\section{Quantum mechanics of two times \label{s2}}There is more than one approach to extend wave equations to include two time-like axes. The use of augmented metrics has been proposed \cite{Hosotani:1983xw, Bonezzi:2010prd, SS1979, Bars2001, BarsT2010, BarsK1997B, Bars2008, BarsDM1999, VNRS2005, BarsK1997D}, which can be traced back to the least action principle where densities are defined in an extended Minkowski space. 

In connection with the many body problem in relativity, we have already discussed some similarities and differences with our topic. It should be stressed now that, although such theories exist, the triumphant view of our physical world is the one provided by Relativistic Quantum Field Theory \cite{ryder}. As explained in many textbooks, only one set of coordinates is needed and many-particle states are represented in second quantization with Fock states. One could reproduce here the first attempts to describe Quantum Electrodynamics using multiple sets of coordinates for $N$ charged fermions with $N$ arbitrary but fixed. It is interesting that the first physical theory that necessitated multiple times was proposed by the old masters of quantum theory, precisely in this direction: Dirac \cite{Dirac2}, Bloch \cite{bloch} and Tomonaga \cite{tomonaga}. Additional efforts in the study of multi-particle physics with many times were made in more recent years, e.g. \cite{lienert1, crater, sazdjian}. 

Our quantum mechanical study involves two lines of reasoning that are different, in principle. In our first approach we shall explore the behavior of a quantum field in the context of augmented metrics, leading to very stringent conditions on the evolution. In our second approach, we shall derive conditions for arbitrary Hamiltonians and states, without alluding to coordinates at the level of wavefunctions. This treament is an extension of our previous premises for the classical case: two dynamical non-preferential equations. 

We shall see a clear physical difference between the two approaches, the second being the most general. Augmented metrics imply extensions of a single wave equation, while unitary evolution of two times demands the existence of two Schr\"odinger equations. As a consequence, the second approach is more permissive, providing less restrictive conditions for the possibility of including the variation of physical quantities in two times.

In any case, the general meaning of a wavefunction is not affected by our considerations; let us spell out its interpretation in the context of two times. If $|\psi\>$ is a physical state and $|x\>$ represents a state of coordinates $x$, then the probability amplitude of finding a physical system\footnote{In the case where wave equations are extended to two times, the coordinates $x$ are those of a particle. The coordinates could also describe the position of many particles in section \ref{s2.2}, but these coordinates could describe as well the position in configuration space of many other physical systems such as the Euler angles of a rotor, the displacement of an elastic string with respect to equillibrium positions, and so on.} between $x$ and $x + dx$ is simply $\<x|\psi\>$. At later times, because of the superposition principle --we have a Hilbert space-- there is a linear operator that transforms the state $|\psi\>$ to $|\psi (t_1, t_2)\>$, dropping the initial times, and therefore the amplitude evolves to $\<x|\psi(t_1,t_2)\>$ -- note that the evolution operator may arise from a prescribed differential equation, such as Klein-Gordon, Dirac, etc. The density $\rho$ is thus the probability of finding the physical system in the coordinates between $x$ and $x+dx$ at the values $t_1, t_2$ of the time parameters, and it is given by $\rho = |\<x|\psi(t_1,t_2)\>|^2$. With appropriate modifications, this is even the case for Klein-Gordon equations --second order in time-- as shown, for example, in \cite{mostafazadeh2006}.

\subsection{Extended wave equations \label{s2.1}}

Quantum physics rests on the assumption that wavefunctions and probabilites are related. In any theory of dimensionally extended wave equations, it is compulsory to include a conserved quantity that can be identified with total probability (with the proper normalization $P=1$). As in any field theory, the alluded conservation law is ensured by N\"other's theorem and therefore by the presence of symmetry. Taking this as a starting point, leads us to consider field equations with gauge symmetries, such that quantum-mechanical waves describing the evolution of matter can be related to a continuity equation satisfied by the corresponding N\"other currents. A specific example with the $1+2$ dimensional Dirac equation is discussed in appendix \ref{app3}. In the general case, we only need to analyze the continuity equation for extended dimensionality.

The idea is to find a conserved probability along two times, avoiding the case in which this occurs only for one time. The latter obviously implies that there can be only one-time dynamics, while the former still makes a case for non-trivial physics. Later, we shall find that in this general scenario, the probability density must be very special. We start with the extended continuity equation

\bea
\partial_{\mu} j^{\mu} &=& 0, \quad g_{\mu \nu} = \mbox{diag}\left\{+,+,-,\cdots,- \right\}, \nonumber \\ \mu, \nu &=& 1,\cdots, d+2. 
\label{2.1}
\eea
In order to obtain the conserved 'charge' (in the N\"other sense), we must split this equation into time-like and spce-like components, with the aim of performing space integration under some specific boundary conditions. In our previous notation, we write

\bea
\nabla_{12} \cdot \v j_{\mbox{\scriptsize time}} 
- \nabla_{\v x} \cdot \v j_{\mbox{\scriptsize space}}
= 0,
\label{2.2}
\eea
with
\bea
\v j_{\mbox{\scriptsize time}} &=& (j_1,j_2), \nonumber \\ \v j_{\mbox{\scriptsize space}} &=& (j_3,\cdots, j_{d+2})=(j_x,j_y, j_z, \cdots ), \\
\nabla_{12} &=& (\partial_1,\partial_2), \nonumber \\ \nabla_{\v x} &=& \left( \frac{\partial}{\partial x^1},\cdots, \frac{\partial}{\partial x^d} \right).
\label{2.3}
\eea
Now we are ready to integrate over a $d+1$ dimensional space in order to obtain the total derivative of some quantity; this can be done in two different ways, over volumes $V_1$ and $V_2$:

\bea
\frac{d}{d t_1} \left\{  \int_{V_1} dt_2 d\v x j_1 \right\} = \int_{V_1} dt_2 d \v x \left( \partial_2, \nabla_{\v x} \right) \cdot \left( -j_2, \v j_{\mbox{\scriptsize space}} \right) \nonumber \\ 
\label{2.4}
\eea

\bea
\frac{d}{d t_2} \left\{  \int_{V_2} dt_1 d\v x j_2 \right\} = \int_{V_2} dt_1 d \v x \left( \partial_1, \nabla_{\v x} \right) \cdot \left( -j_1, \v j_{\mbox{\scriptsize space}} \right). \nonumber \\
\label{2.5}
\eea
but now we find two conserved quantities after applying the divergence theorem to the r.h.s. and imposing vanishing integrals on the boundary (which is always necessary, even in single-time dynamics)\footnote{Of course, a drop-off condition for bound states implies a vanishing surface integral (trivially), but scattering states with real energies also comply with zero surface integrals, since the total incoming flux is equal to the outgoing flux. The latter indicate that our extended wavefunctions do not necessarily vanish for $t_i \rightarrow \pm \infty$, i.e. no localization in time is required.}. This leads to

\bea
\frac{d Q_i(t_i)}{dt_i} &=& \oint_{S_i} d \v s \cdot  \left( -j_k, \v j_{\mbox{\scriptsize space}} \right) = 0, \nonumber \\  i&=&1,2, \quad  k=1,2, \quad k \neq i \nonumber \\
Q_i (t_i) &\equiv&  \int_{V_i} dt_k d\v x \, j_i \qquad \mbox{Conserved charge.}
\label{2.6}
\eea
From here we see that there are two densities $\rho_i = \int dt_k j_i$ corresponding to each conserved charge. Our task now is to find a conserved quantity in both times $Q(t_1,t_2)$ using (\ref{2.6}). As a consequence, we shall see that any probability density such that

\bea
\int d \v x \rho(\v x; t_1, t_2) = Q(t_1,t_2) = 1
\label{2.7}
\eea
must be a separable function of $t_1,t_2$. In order to obtain an expression for $\rho$ in terms of the aforementioned densities $\rho_i$, let us consider a functional $\rho \left[ \rho_1, \rho_2 \right]$ and replace it back in (\ref{2.7}). Computing the time derivatives

\bea
\partial_i \int d \v x \rho \left[ \rho_1, \rho_2 \right] &=& \partial_i Q(t_1,t_2) = 0 \nonumber \\
&=& \int d \v x \left\{ \frac{\partial \rho}{\partial \rho_1} \partial_i \rho_1 + \frac{\partial \rho}{\partial \rho_2} \partial_i \rho_2 \right\} \nonumber \\
&=&   \int d \v x \frac{\partial \rho}{\partial \rho_i} \partial_i \rho_i\qquad 
\label{2.8}
\eea
where the last line follows from $\partial_1 \rho_2 = \partial_2 \rho_1 =0$. Moreover, since this integral vanishes, the result must be independent of $t_1$ and $t_2$. If it is independent of $t_1$ for $i=2$, then $\partial \rho / \partial \rho_2$ cannot depend on $\rho_1$ (unless $\rho_1$ is trivially a constant) and a similar argument yields that $\partial \rho / \partial \rho_1$ cannot depend on $\rho_2$. This means that $\rho$ must be separable $\rho = \alpha g_1(\rho_1) + \beta g_2(\rho_2)$ with $\alpha, \beta$ two constants. Now we consider $i=2$ when the integral does not depend on $t_2$, which leads to 

\bea
\int d \v x \beta g_2'(\rho_2) \partial_2 \rho_2 = 0 = \partial_2 Q_2 = \int d \v x \partial_2 \rho_2,
\label{2.9}
\eea  
and finally the desired result is $\beta g_2' = 1$, i.e. $\rho$ is linear in $\rho_1, \rho_2$. This is indeed a very strong result:

\bea
\rho = \alpha \rho_1 + \beta \rho_2.
\label{2.10}
\eea 
As a consequence, the total charge is also separable $Q = \alpha Q_1 + \beta Q_2$, and the constants $\alpha, \beta$ can be chosen in terms of the normalization condition.
We finish our discussion by recognizing that separability has a restrictive effect on averages, when computed with our $\rho$. The average of any function of the position (say $A(\v x)$) is separable as well

\bea
\< A \> = \int d \v x A \rho_1 + \int d \v x A \rho_2.
\label{2.11}
\eea
A dynamical consequence of this property is that $\<\v x \>$ itself is separable, i.e. $\< x^i \> = f^i_1(t_1) + f^i_2(t_2)$. According to the Ehrenfest's theorem, in the classical limit (for any {\it bona fide\ }quantization scheme) one must recover Hamilton's  equations of motion for these averages. The force (as treated in our previous discussions) should act then in such a way that the evolution is separable, leading to

\bea
F_{12}^i = \partial_1 \partial_2 \< x^i \> = 0
\label{2.12}
\eea
but in general we have a non-vanishing diagonal
\bea
F_{11}^i = \partial_1^2 \< x^i \>\neq 0, \quad F_{22}^i = \partial_2^2 \< x^i \>\neq 0. 
\label{2.13}
\eea
When the dynamical equations (\ref{2.13}) or their classical counterparts (\ref{2}) are decoupled in this way, one finds a strong consequence:

\bea
F_{jj}^i ( \< \v x \> ) &=&  F_{jj}^i ( \v f_1(t_1) + \v f_2(t_2)  ) = \partial_j^2 \left[f^i_1(t_1) + f^i_2(t_2) \right] \nonumber \\ &=&  \partial_j^2 f^i_j(t_j) 
\label{2.14}
\eea
so now $F^i_{11}$ may depend only on $t_1$ and $F^i_{22}$ may depend only on $t_2$. For this reason, either $f^i_1 =$ constant or $f^i_2=$ constant. In this way we have shown that in the classical limit, this scheme forces a single time dependence of the average position. The external forces are also constrained, in the sense that any external stimulus must produce evolution in only one parameter. This argument is valid for $i=1,\cdots,d$, but it is possible to analyze simple cases such as $d=1$, where the force tensor must have only one non-vanishing component.

Once more, parallels can be drawn regarding multi-particle theories. Tensor conservation laws have been formulated with the aim of including every particle current in a vanishing integral \cite{sazdjian}. It seems that this approach has been very helpful in ensuring conservation of probabilities where times are necessarily attached to particle labels. However, such treatments do not include two times for a single particle -- they do not reduce to such a case. A vector current (as opposed to tensor) for a single particle in $d+2$ dimensions obeys a different conservation law, since a partition of the remaining $d$ spatial coordinates in 3d spaces is not necessary, nor the imposition of a vanishing integral of a tensor for surfaces covering each particle's coordinates. With this, we conclude that the separability results of the density in this section are quite restrictive. 

\subsection{Unitary evolution with two times \label{s2.2}}

 It is important to stress that the issue of probability conservation has been raised regarding extended differential equations using the first approach. For this reason, in our second approach we postulate a Hilbert space of physical states $|a\>$ evolving under two parameters by means of a unitary transformation $U(t_1,t_2)$ such that

\bea
U(t_1,t_2; t_{01}, t_{02}) |a\> = |a(t_1,t_2)\>, \quad UU^{\dagger}=U^{\dagger}U=1. \nonumber \\
\label{12}
\eea  
Let us drop the initial time label $ t_{01}, t_{02}$ to simplify the notation. An infinitesimal expansion of $U$ reveals a Hermitean generator $H(t_1,t_2)$ which is in fact composed by two operators in each time direction $H_1, H_2$. By a standard procedure, we can show that each evolution leads to a Schr\"odinger equation for the operator $U$ in terms of the corresponding generator (note that this is deduced and not postulated):

\bea 
H_i(t_1,t_2) U(t_1,t_2) = i \hbar \partial_i U(t_1,t_2), \qquad i=1,2.
\label{13}
\eea
This is equivalent to a unitary group with two parameters, but in order to elucidate its structure we must resort once more to smoothness conditions on $U$, at least in some region of the space $t_1,t_2$. Commutability of derivatives implies

\bea
\partial_1(H_2 U) = \partial_2 (H_1 U),
\label{14}
\eea
or multiplying by $U^{\dagger}$ from the right and using (\ref{13})

\bea
i \hbar \partial_{\left[ i \right.} H_{\left. j \right]} =  \left[H_i, H_j \right],
\label{15}
\eea
which tell us that the generators obey fundamental non-commutability conditions in terms of their derivatives. Finally, (\ref{13}) can be used to establish two Schr\"odinger equations when acting on a specific state. Similar conditions were derived in \cite{petrat}. In order to work out that the result is, in fact, independent of any interpretational relation to multi-particle physics, as well as to show that the multiple times do not necessarily need to be attached to particles, we have presented the result in a different way.

In addition, we would like to mimic here the treatment used in the classical case by analyzing the evolution of observables. The Heisenberg picture can be extended naturally in this framework if $A^{\mbox{\scriptsize H}}\equiv U^{\dagger}AU$, leading to

\bea
i \hbar \partial_i A^{\mbox{\scriptsize H}} = \left[ A^{\mbox{\scriptsize H}}, H^{\mbox{\scriptsize H}}_{i} \right], \quad i=1,2.
\label{16}
\eea
For any observable, in particular for position $x$, we have again two velocities \footnote{Here we do not need to impose any representation on momenta, nor a canonical commutator with $x$. However, note that the prescription $p=-i\hbar \partial/\partial x$ must be modified to accomodate two times.}:

\bea
v_i = - \frac{i}{\hbar} \left[ x, H_i \right]
\label{17}
\eea
and four possible accelerations:

\bea
a_{ij} = \partial_i v_j = - \frac{1}{\hbar^2} \left[ \left[ x, H_j \right], H_i \right] \equiv F_{ij}.
\label{18}
\eea
This is in fact Newton's second law if we admit that the dynamics is governed by previous specification of the operator $F_{ij}$ \footnote{ Equations of a finite order in time are sufficient. Quantum mechanical versions of the Abraham-Lorentz force could be proposed, but this would make our treatment lengthier.}. With these relations we may now establish a set of dynamical equations for averages using a fixed state $|\psi\>$:

\bea
\< a_{ij} \> = \partial^2_{ij}\<x\> =\<F_{ij}\>.
\label{19}
\eea
This is our two-time version of Ehrenfest's theorem; a good classical limit is ensured when $x_{\mbox{\scriptsize Classical}}=\<x\>$ and $F_{ij}(x_{\mbox{\scriptsize Classical}}) = \<F_{ij}\>$. The second condition clearly depends on the state $|\psi\>$ we choose, but imposing such a limit leads directly to our previously obtained conditions, i.e. the evolution of averages must pick one direction in the plane $t_1,t_2$.

\subsection{Fluctuations and fundamental constraints \label{S2.3}} In a full quantum-mechanical regime, we may analyze the possibility of 'leakage' of the evolution along other directions. In a general framework, a reasonable assumption comes into play: If physics is to be invariant under translations of the origin of time, then $\partial_i H_j =0$ and consequently $\left[H_i,H_j\right]=0$  \footnote{In the general case $\partial_i H_j \neq 0$, we can prove that smoothness of $x(t_1,t_2)$ leads in fact to $\left[x, \left[H_i, H_j \right] \right]=0$, i.e. $i \left[H_i, H_j \right]$ must be compatible with $x$ and it may contain other physical observables such as spin, color, flavor, etc.}. Therefore, the Hamiltonian generators share an eigenbasis and we may analyze the dynamics in a suitable matrix representation where $x^{mn}=\<m|x|n\>, H_i |n\>=E_i^n |n\>$. Computing the matrix elements of both sides of (\ref{18}) and defining $\Delta^{nm}_i \equiv E^n_i-E^m_i$ yields 

\bea
\partial_i \partial_j x^{nm}= x^{nm}\Delta^{nm}_i \Delta^{nm}_j,
\label{20}
\eea
which is symmetric in $i,j$, as expected. Once more, velocities $v_i$ and their matrix elements $v_i^{nm}=i\Delta^{nm}_i x^{nm}/\hbar$ can be assumed to be smooth functions, rendering the following conditions

\bea
\Delta^{nm}_1 \partial_2 x^{nm} - \Delta^{nm}_2\partial_1 x^{nm} = 0.
\label{21}
\eea
The evolution of each matrix element is thus restricted by a geometric condition of the type

\bea
\v \fcal^{nm} \cdot \nabla_{12} \, x^{nm}=0 
\label{22}
\eea
where the field $\v \fcal^{nm}=(\Delta^{nm}_2,-\Delta^{nm}_1)$ now depends on energy quanta $n,m$, in contrast with its classical counterpart (\ref{11}). A convenient change of variables in the two-time plane helps us to understand the nature of each direction; we propose a rotation of variable angle

\bea
\tau^{nm}_1 &=& \cos (\theta^{nm}) t_1 + \sin( \theta^{nm}) t_2 \nonumber \\
\tau^{nm}_2 &=& -\sin(\theta^{nm}) t_1 + \cos(\theta^{nm}) t_2
\label{23}
\eea
with
\bea
\cos (\theta^{nm}) &=& \frac{\Delta^{nm}_1}{\sqrt{(\Delta^{nm}_1)^2+(\Delta^{nm}_2)^2}}=\frac{\Delta^{nm}_1}{|\fcal^{nm}|}, \nonumber \\
\sin (\theta^{nm}) &=& \frac{\Delta^{nm}_2}{\sqrt{(\Delta^{nm}_1)^2+(\Delta^{nm}_2)^2}}=\frac{\Delta^{nm}_2}{|\fcal^{nm}|}.
\label{24}
\eea
The geometric restriction (\ref{22}) now implies that $x^{nm}$ depends only on the first variable $\tau^{nm}_1$ and not on $\tau^{nm}_2$. In other words, each element 'chooses' its own time direction as a function of the energies. When $n=m$ we recover the classical limit, but in the full quantum picture the single-time evolution is applicable only to individual processes. The possibility of having an overall two-time physics may take place, provided that fluctuations in $x$ or any other observable are non-vanishing:

\bea
\sigma^2 &=&\<x\>_{\psi}^2-\<x^2\>_{\psi} \nonumber \\ 
&=& x_{\mbox{\scriptsize Classical}}^2 - \sum_{nm} \psi_m^* \psi_n |x^{nm}(0)|^2 \exp\left(\frac{i \tau^{nm}_1 |\fcal^{nm}|}{\hbar}\right) \nonumber \\
\label{25}
\eea
where $\psi_n = \<n|\psi\>$ and $\<x\>$ is now replaced by its limit $x_{\mbox{\scriptsize Classical}}$. The average $\<x^2\>$ is what we must analyze in our search for residual effects. Although many elaborate procedures can be employed in the study of quantum evolution and decoherence --e.g. density operators and master equations \cite{Zurek:2003, Schlosshauer:2005, Isar:1994}-- the expression in (\ref{25}) serves well our purposes; the more terms involved in $\<x^2\>$, the more plausible is to have superpositions along many time directions $\tau_1^{nm}$. Quantumness thus appears as a bound to such many possible choices of single-time dynamics. A second effect deserves a mention: the effective time of the evolution becomes one and only one when the angle is fixed $\theta^{nm}=\theta$, which is tantamount to saying that a single direction has been picked by the system rather than the elimination of $t_2$ in favor of $t_1$. A possible scenario occurs when a proportionality $E_1^{n}-E_1^{m} = \epsilon (E_2^{n}-E_2^{m}) $ holds, i.e. when the generators coincide in their shape $H_1 = \epsilon H_2 + E_0$. In our attempt to explore new physics, it is worth noting that the ressemblance between the spectra of $H_1$ and $H_2$ leads to a small probability of finding two-time dynamics in the fluctuations of any observable.

\subsection{Extended uncertainty principle \label{S2.3}}As a continuation of our analysis, we note now that (\ref{25}) would contain strong oscillations -- therefore vanishing contributions -- unless the argument of the exponential satisfies a new uncertainty relation:

\bea
(E^n_1-E^m_1) t_1 + (E^n_2-E^m_2) t_2 \sim \hbar
\label{26}
\eea
or in compact notation

\bea
\Delta E_1 \Delta t_1 + \Delta E_2 \Delta t_2 \sim \hbar.
\label{27}
\eea
The inequality is also attainable, but if the argument of the exponential is too small, there will be no chance to perform at least one oscillation and the evolution would not have produced significant variations on $\<x^2\>$. 

As a final task, let us estimate the variation in the angle of the trajectories around some average value. We define a length in time variables $t = \sqrt{t_1^2+t_2^2}$ and take $\phi$ as the angle between the vectors $(t_1,t_2)$ and $(\Delta_1^{nm},\Delta_2^{nm})=(\Delta E_1, \Delta E_2)$. We have, by virtue of (\ref{27})

\bea
\cos \phi = \frac{\hbar}{t \sqrt{(\Delta E_1)^2 + (\Delta E_2)^2}} \leq 1.
\label{28}
\eea
The effects in quantum fluctuations, as we have seen, come from variations of energies as functions of $n,m$. In order to estimate the fluctuations on $\phi$, we obtain the differential $\delta \phi$ by computing the derivative of $\cos \phi$ in both sides of (\ref{28}). This allows to express $\delta \phi$ in terms of increments of energy spacings $\delta(\Delta E_i)$: 

\bea
\delta \phi = \frac{\hbar \left[\Delta E_1 \delta(\Delta E_1)  + \Delta E_2 \delta(\Delta E_2)   \right]}{\left[ (\Delta E_1)^2 + (\Delta E_2)^2 \right] \sqrt{t^2\left[ (\Delta E_1)^2 + (\Delta E_2)^2 \right]-\hbar^2}}, \nonumber \\
\label{29}
\eea
and if $\hbar$ is retained to lowest order, we have the lowest possible estimate:

\bea
\delta \phi \sim \frac{\hbar \left[\Delta E_1 \delta(\Delta E_1)  + \Delta E_2 \delta(\Delta E_2)   \right]}{t \left[ (\Delta E_1)^2 + (\Delta E_2)^2 \right]^{3/2} }.
\label{30}
\eea
Our chances to detect this small width $\delta \phi$ from our new uncertainty principle are greatly bound by level spacings of both generators of the evolution. The inequality in (\ref{28}) can be applied to find

\bea
\delta \phi \leq \frac{\Delta E_1 \delta(\Delta E_1)  + \Delta E_2 \delta(\Delta E_2) }{ (\Delta E_1)^2 + (\Delta E_2)^2  },
\label{31}
\eea
which is now independent of physical constants. Although this bound does not guarantee the existence of two time axes, we can be sure that the particular details of Hamiltonians $H_1,H_2$ -- in other words, the model of our physical system -- are enough to constrain observability. From the denominator of  (\ref{31}) we can see that large spacings destroy any possibility of a finite width. We must stress though that the inequality is controlled also by the fluctuations $\delta(\Delta E)$, alluding now to the composition of wavepackets comprised of many energy scales. The energy differences can indeed fluctuate if $|\psi\>$ contains both a fine and a coarse structure of levels. We have reached thus the conclusion that the richness in composition as a whole gives the opportunity of observing sustained two-time evolution. A two-level system --the most quantum mechanical example that we know-- does not suffice, nor an experiment of very massive particles isolated in the scale $100$ GeV. 

\section{Conclusion \label{s3}} The introduction of two times through non-preferential and probability-conserving laws leads to strong bounds. We have seen how quantum mechanics may be less restrictive, where the roles of Planck's constant and the level spacing fluctuations are equally important. Violations to our fundamental constraints would imply severe deviations from known physics, such as an invalid Newton's second law or higher order differential equations. In the quantum domain, probability conservation is attached to the very notion of probability and we do not believe that such a strong hypothesis could be dropped.



\begin{acknowledgement}
Financial support from CONACyT under project CB 2012-180585 is acknowledged. 
\end{acknowledgement}

\appendix

\section{Parallel vector fields}

\subsection{Fields in 2+2 dimensions \label{app1}}

The components of the fields that restrict the dynamics for the $2+2$ dimensional case are

\bea
\ccal_1 = \begin{array}{|cc|}\tilde \alpha&\tilde \beta\\ \tilde \gamma& \tilde \delta\end{array} , \ccal_2 = \begin{array}{|cc|}\tilde \alpha'&\tilde \beta'\\ \tilde \gamma'& \tilde \delta'\end{array} , \\
\dcal_1 = \begin{array}{|cc|}\alpha&\beta\\ \gamma&\delta\end{array}, \dcal_2= \begin{array}{|cc|}\alpha'&\beta'\\ \gamma'&\delta'\end{array},
\label{a1.7}
\eea
with the quantities 
\bea
\alpha=\begin{array}{|cc|}F_{21,x}^x&F_{21,y}^x\\ F_{21,x}^y&F_{21,y}^y\end{array},\beta=\begin{array}{|cc|}F_{11,x}^x&F_{12,x}^x\\ F_{11,x}^y&F_{21,x}^y\end{array}, \nonumber \\ \gamma=\begin{array}{|cc|}F_{21,y}^x&F_{22,y}^x\\ F_{21,x}^x&F_{22,x}^x\end{array},\delta=\begin{array}{|cc|}F_{21,x}^x&F_{12,x}^x\\ F_{21,x}^x&F_{22,x}^x\end{array},  \nonumber \\
\label{a1.8}
\eea
\bea
\tilde{\alpha}=-\alpha,\tilde{\beta}=\begin{array}{|cc|}F_{11,y}^x&F_{21,y}^x\\ F_{11,y}^y&F_{21,y}^y\end{array},\tilde{\gamma}=-\gamma,\tilde{\delta}=\begin{array}{|cc|}F_{11,y}^x&F_{12,y}^x\\ F_{21,y}^x&F_{22,y}^x\end{array},
\label{a1.9}
\eea
\bea
\alpha'=\begin{array}{|cc|}F_{21,x}^x&F_{11,y}^x\\F_{21,x}^y&F_{11,y}^y\end{array},\beta'=\beta,\gamma'=\begin{array}{|cc|}F_{11,y}^x&F_{12,y}^x\\F_{21,x}^x&F_{22,x}^x\end{array},\delta'=\delta,
\label{a1.10}
\eea
\bea
\tilde{\alpha'}=\begin{array}{|cc|}F_{21,y}^y&F_{11,x}^y\\ F_{21,y}^x&F_{11,x}^x\end{array},\tilde{\beta'}=\tilde{\beta},\tilde{\gamma'}=\begin{array}{|cc|}F_{22,y}^x&F_{12,x}^x\\F_{21,y}^x&F_{11,x}^x\end{array},\tilde{\delta'}=\tilde{\delta}.
\label{a1.11}
\eea

\subsection{Fields in 3+2 dimensions \label{app2}}

The smoothness of the functions $p^{i}_j$ and the chain rule applied to the derivative of the force lead, in this case, to the following linear system

\begin{equation} \left(\begin{array}{cccccc}
	F_{21,x}^1&-F_{11,x}^1&F_{21,y}^1&-F_{11,y}^1&F_{21,z}^1&-F_{11,z}^1\\
	F_{21,x}^2&-F_{11,x}^2&F_{21,y}^2&-F_{11,y}^2&F_{21,z}^2&-F_{11,z}^2\\
	F_{21,x}^3&-F_{11,x}^3&F_{21,y}^3&-F_{11,y}^3&F_{21,z}^3&-F_{11,z}^3\\
	F_{22,x}^1&-F_{12,x}^1&F_{22,y}^1&-F_{12,y}^1&F_{22,z}^1&-F_{12,z}^1\\
	F_{22,x}^2&-F_{12,x}^2&F_{22,y}^2&-F_{12,y}^2&F_{22,z}^2&-F_{12,z}^2\\
	F_{22,x}^3&-F_{12,x}^3&F_{22,y}^3&-F_{12,y}^3&F_{22,z}^3&-F_{12,z}^3\end{array}\right) \left(\begin{array}{c} p^{1}_1 \\  p^{1}_2 \\  p^{2}_1 \\ p^{2}_2 \\  p^{3}_1 \\ p^{3}_2  \end{array} \right)=0.
\label{a1.12}
\end{equation}
From here, and the condition of a vanishing determinant (\ref{1.9}), we may eliminate successively the variables $p_2^3$ in terms of other $p'$s, $p_1^3$ in terms of the remaining $p'$s and so forth. All the resulting relations are linear in $p$, but not in $F$. The final substitution is equivalent to the first condition in (\ref{1.10}): 

\bea
\ccal_{1,1} p_1^1 + \ccal_{1,2} p_2^1 =0 = \ccal_1 \cdot \nabla_{12} x.
\label{a1.13}
\eea
Once this expression has been established, we may solve for other $p$'s in order to obtain the remaining two

\bea
\ccal_{2,1} p_1^2 + \ccal_{2,2} p_2^2 =0 = \ccal_2 \cdot \nabla_{12} y, \nonumber \\
\ccal_{3,1} p_1^3 + \ccal_{3,2} p_2^3 =0 = \ccal_1 \cdot \nabla_{12} z,
\label{a1.14}
\eea
where the components of the three vectors $\ccal_i, i=1,2,3$ can be solved completely in terms of $F_{mn,k}^{i}$. This completes the procedure, but as an example we provide $\ccal_1$ explicitly (the other three can be obtained by trivial permutations):

\begin{equation}
\ccal_{1,j} =	\det\left(\begin{array}{cc}
		A^3_2(j)&A^3_2(3)\\
		A^3_3(j)&A^3_3(3)\end{array}\right)		,\qquad j=1,2
\end{equation}
where

\begin{equation}
A^3_m(n)=\det\left(\begin{array}{cc}
	A^2_m(n)&A^2_m(4)\\
	A^2_4(n)&A^2_4(4)
\end{array}\right), \qquad m,n = 1, \cdots, 3.	
\end{equation}

\begin{equation}
A^2_m(n)=\det\left(\begin{array}{cc}
	A_m^1(n)&A^1_m(5)\\
	A^1_5(n)&A^1_5(5)
\end{array}\right), \qquad m,n = 1, \cdots, 4.
\end{equation}

\begin{equation}
A^1_m(n)=\det\left(\begin{array}{cc}
	c_{mn}&c_{m1}\\
	c_{6n}&c_{61}
\end{array}\right), \qquad m,n = 1, \cdots, 5.
\end{equation}
and the matrix $c_{mn}$ is given in terms of the force tensor

\begin{equation} 
(c_{mn})=
\left(\begin{array}{cccccc}
	F_{21,x}^1&-F_{11,x}^1&F_{21,y}^1&-F_{11,y}^1&F_{21,z}^1&-F_{11,z}^1\\
	F_{21,x}^2&-F_{11,x}^2&F_{21,y}^2&-F_{11,y}^2&F_{21,z}^2&-F_{11,z}^2\\
	F_{21,x}^3&-F_{11,x}^3&F_{21,y}^3&-F_{11,y}^3&F_{21,z}^3&-F_{11,z}^3\\
	F_{22,x}^1&-F_{12,x}^1&F_{22,y}^1&-F_{12,y}^1&F_{22,z}^1&-F_{12,z}^1\\
	F_{22,x}^2&-F_{12,x}^2&F_{22,y}^2&-F_{12,y}^2&F_{22,z}^2&-F_{12,z}^2\\
	F_{22,x}^3&-F_{12,x}^3&F_{22,y}^3&-F_{12,y}^3&F_{22,z}^3&-F_{12,z}^3\end{array}\right).
	\end{equation}

\section{Continuity in the $1+2$ Dirac equation}
\label{app3}
The usual $d+1$ dimensional Dirac equation has an associated conserved current which is well understood. We proceed to extend the Dirac equation to $1+2$ dimensions and obtain an appropiate conserved current, and therefore a suitable probability density. We find that the derived conserved current is consistent with section \ref{s2.1}. Thus, we end up with the same conclusions, and we have a classical limit with a single time evolution.

With the energy-momentum conservation equation ($\hbar=c=1$)

\begin{eqnarray}
	\left\{\Box_{1+2}+m^2\right\}\Psi=0
\label{A2.1}
\end{eqnarray}

as the starting point, we want to obtain an equation of order 1 in the two times. In order to do so, a set of Dirac $\gamma$ matrices is needed. Denoting time components by $\mu = 1,2$ and the space component by $\mu=3$, we write the square root of (\ref{A2.1}) as

		\begin{eqnarray}
\left\{i\gamma_\mu\partial^\mu-m\right\}\Psi(\v x)=0,
\label{A2.2}
\end{eqnarray}
where we use Clifford's condition

	\begin{eqnarray}
\left\{\gamma_\mu,\gamma_\nu\right\}=2g_{\mu\nu} \v 1,
\label{A2.3}
\end{eqnarray}
and $g_{\mu\nu}$ is given by (\ref{2.1}); we find that a $2\times 2$ representation is enough, and the $\gamma$ matrices may be given by

	\begin{eqnarray}
\gamma_3 = \gamma_x=i\sigma_3,\quad
\gamma_1=\sigma_1,\quad \gamma_2=\sigma_2.
\label{A2.4}
\end{eqnarray}
With the extended Dirac equation (\ref{A2.2}), now we proceed to find a conserved current $j^\mu$,

	\begin{eqnarray}
\partial_\mu j^\mu=0,
\label{A2.5}
\end{eqnarray}
ensured by N\"oether's theorem. 
The usual way (in $3+1$ dimensions) to find such a conserved current does not work well in this case, as a consequence of the effective hamiltonian not being hermitean:

	\begin{eqnarray}
i\partial_1\Psi=-i\partial_2[(\gamma_1\gamma_2)\Psi]+i\frac{\partial}{\partial x}[(\gamma_1\gamma_3)\Psi]+m\gamma_1\Psi.
\label{A2.6}
\end{eqnarray}
This further implies that the field describing the antiparticle cannot be defined as $\Psi^\dagger\gamma_3$. 
However, by making a coordinate inversion while considering the hermitean conjugate of Dirac equation

	\begin{eqnarray}
\Psi^\dagger(-x_\nu)\left\{+i\gamma^\dagger_\mu\overleftarrow{\partial}^\mu-m\right\}=0,
\label{A2.7}
\end{eqnarray}
we are led to a conservation equation of the form

	\begin{eqnarray}
\partial^\mu\left[i\Psi^\dagger(-x_\nu)\gamma_3\gamma_\mu\Psi(x_\nu)\right]=0.
\label{A2.8}
\end{eqnarray}
For the current, we consider only the hermitean part of the previous quantity

	\begin{eqnarray}
j_\mu=\frac{1}{2}i\Psi^\dagger(-x_\nu)\gamma_3\gamma_\mu\Psi(x_\nu)+ \mbox{h.c.}
\label{A2.9}
\end{eqnarray}
and the $\bar{\Psi}$ describing antiparticles must be defined as

	\begin{eqnarray}
\bar{\Psi}(x_\nu)=\Psi^\dagger(-x_\nu)\gamma_3.
\label{A2.10}
\end{eqnarray}
With this conserved current we can obtain the probability density and total charge as done in section \ref{s2.1}:

	\begin{eqnarray}
\rho({\bf x} ;t_1,t_2)=\alpha\int dt_2j_1+\beta\int dt_1j_2 \nonumber \\ =\alpha\rho_1({\bf x},t_1)+\beta\rho_2({\bf x},t_2)
\label{A2.11}
\end{eqnarray}
and

	\begin{eqnarray}
P=\int d{\bf x}\rho({\bf x};t_1,t_2)=\alpha P_1(t_1)+\beta P_2(t_2);
\label{A2.12}
\end{eqnarray}
 which, once again, are separable. The conclusions in section \ref{s2.1} apply, and then, in the classical limit, the average position evolves as a function of only one time.

\subsection{Positivity of densities}

Probability densities must be positive quantities in all space. Although the Klein-Gordon current corresponding to (\ref{A2.1}) could be used as a conserved current, textbook observations on such currents reveal that only positive energy components can be related to positive densities; the conserved quantity here is the {\it charge.\ } For Dirac currents, it is left to prove that the $1+2$ theory allows positive densities. If we go back to (\ref{A2.11}), we will note that the positivity is a property that corresponds to $\rho$ in general. In particular, if $j_1$ and $j_2$ are positive, then the result should follow, but these are not the most general conditions. It seems appropriate, though, to find the restrictions in the solutions of the two-time Dirac's equation that allow such a scenario. Compare with \cite{Lienert2015} and the structure of the resulting theory.

We start by noting that either the imaginary part or the real part of (\ref{A2.9}) can be used as a valid current. As long as the sign of these components does not change, we may define $j_1,j_2$ up to a sign. In order to compute $j$'s, we write the solutions of (\ref{A2.2}) as linear combinations of on-shell waves

\bea
\Psi = e^{i k_{\mu} x^{\mu}} \Psi_{+}(0) + e^{-i k_{\mu} x^{\mu}} \Psi_{-}(0).
\label{app2.1.1}
\eea
Substitution of this Dirac spinor on Dirac's equation yields the conditions

\bea
\left\{-\gamma_{\mu} k^{\mu} - m \right\} \Psi_{+}(0) = 0,
\label{app2.1.2}
\eea

\bea
\left\{\gamma_{\mu} k^{\mu} - m \right\} \Psi_{-}(0) = 0,
\label{app2.1.3}
\eea
while the vanishing determinant of each linear operator is simply the on-shell condition, which is independent of the sign of $k_{\mu}$. Now, in terms of components, the spinors are

\bea
\Psi_{\pm}(0) = \left( \begin{array}{c} C_{\pm}^1 \\ C_{\pm}^2 \end{array} \right)
\label{app2.1.4}
\eea
and substitution of (\ref{app2.1.4}) in the imaginary part $\Im \left[\cdot \right]$ of (\ref{A2.9}) gives 

\bea
j_1 = 2 \cos(2k_{\mu}x^{\mu}) \Im\left[ C_{+}^{2 *} C_{+}^{1} + C_{-}^{2 *} C_{-}^{1}  \right] \nonumber \\ +
2 \Im\left[ C_{-}^{2 *} C_{+}^{1} + C_{+}^{2 *} C_{-}^{1}  \right],
\label{app2.1.5}
\eea

\bea
j_2 = 2 \cos(2k_{\mu}x^{\mu}) \Im\left[ C_{+}^{2 *} C_{+}^{1} + C_{-}^{2 *} C_{-}^{1}  \right] \nonumber \\ +
2 \Im\left[ C_{-}^{2 *} C_{+}^{1} + C_{+}^{2 *} C_{-}^{1}  \right].
\label{app2.1.6}
\eea
Had we used the real part $\Re \left[\cdot \right]$ of (\ref{A2.9}), an overall factor $\sin(2k_{\mu}x^{\mu})$ would have spoilt the result, since this factor changes its sign in space and time. Now, (\ref{app2.1.5}) and (\ref{app2.1.6}) preserve their sign (positive) if the following conditions are met

\bea
|\Im\left[ C_{+}^{2 *} C_{+}^{1} + C_{-}^{2 *} C_{-}^{1}  \right]| \leq
|\Im\left[ C_{-}^{2 *} C_{+}^{1} + C_{+}^{2 *} C_{-}^{1}  \right]|,
\label{app2.1.7}
\eea

\bea
|\Re\left[ C_{+}^{2 *} C_{+}^{1} + C_{-}^{2 *} C_{-}^{1}  \right]| \leq
|\Re\left[ C_{-}^{2 *} C_{+}^{1} + C_{+}^{2 *} C_{-}^{1}  \right]|.
\label{app2.1.8}
\eea
Since (\ref{app2.1.2}) and (\ref{app2.1.3}) relate $C_{\pm}^{1}$ with $C_{\pm}^{2}$, the positivity conditions (\ref{app2.1.7}) and (\ref{app2.1.8}) constrain the normalizations of $\Psi_{+}(0)$, $\Psi_{-}(0)$, together with their relative phase. With these conditions, we have reached positivity. Other possibilities would include more independent solutions using all possible signs in the components of $k_{\mu}$.

\end{document}